\documentclass[a4paper,fleqn,usenatbib]{mnras}

\usepackage{newtxtext,newtxmath}

\usepackage[T1]{fontenc}
\usepackage{ae,aecompl}

\usepackage{graphicx}	
\usepackage{amsmath}	
\usepackage{amssymb}	
\usepackage{siunitx}
\usepackage{lineno,hyperref}

\newcommand{\ca}[1]{{#1}}

\title[Exogenous hydration on Psyche.]{Exogenous origin of hydration on asteroid (16) Psyche: The role of hydrated asteroid families.} 

\author[]{
C. Avdellidou,$^{1}$\thanks{E-mail: chrysa.avdellidou@esa.int}
M. Delbo'$^{2}$,
A. Fienga$^{3}$
\\
$^{1}$Science Support Office, Directorate of Science, European Space Research and Technology Centre (ESA/ESTEC), 2201 AZ Noordwijk,\\ The Netherlands\\
$^{2}$Universit\'e C\^ote d'Azur, CNRS--Lagrange, Observatoire de la C\^ote d'Azur, CS 34229 -- F 06304 NICE Cedex 4, France\\
$^{3}$Universit\'e C\^ote d'Azur, CNRS--Geoazur, Observatoire de la C\^ote d'Azur, 250 Avenue de l' Observatoire -- F 06250 Valbonne, France\\
}

\date{Accepted XXX. Received YYY; in original form ZZZ}

\pubyear{2017}

\begin{document}
\label{firstpage}
\pagerange{\pageref{firstpage}--\pageref{lastpage}}
\maketitle

\begin{abstract}
Asteroid (16) Psyche, that for long was the largest M-type with no detection of hydration features in its spectrum, was recently discovered to have a weak 3~\si{\um} band and thus it was eventually added to the group of hydrated asteroids. Its relatively high density, in combination with the high radar albedo, led to classify the asteroid as a metallic object. It is believed that is possibly a core of a differentiated body, remnant of ``hit-and-run'' collisions. The detection of hydration is, in principle, inconsistent with a pure metallic origin of this body. Here we consider the scenario that the hydration on its surface is exogenous and was delivered by hydrated impactors. We show that impacting asteroids that belong to families whose members have the  3~\si{\um} band can deliver the hydrated material to Psyche.
We developed a collisional model with which we test all the dark carbonaceous asteroid families, which contain hydrated members. We find that the major source of hydrated impactors is the family of Themis, with a total implanted mass on Psyche to be of the order of $\sim$10$^{14}$~kg. However, the hydrated fraction could be only a few per cent of the implanted mass, as the water content in carbonaceous chondrite meteorites, the best analogue for the Themis asteroid family, is typically a few per cent of their mass. 
\end{abstract}

\begin{keywords}
minor planets, asteroids: individual: (16) Psyche
\end{keywords}



\section{Introduction}
\label{introduction}

Throughout the last decades, the study of physical properties of the asteroid (16) Psyche led to the demonstration that it is a peculiar body in our Solar System. It was spectroscopically classified as an M-type \citep{tholen1989}, a taxonomic class containing asteroids with geometric visible albedos between $\sim$0.1 and 0.2 and a generally featureless and slightly red reflectance spectrum, properties being similar to those of the Fe/Ni metal \citep{gaffey2002}. Therefore Psyche, the largest M-type asteroid and target of the NASA space mission \textit{Psyche} \citep{elkins2017}, is considered to be the Fe/Ni core of a differentiated body (or planetesimal) that was exposed by ``hit-and-run'' collisions, capable of removing asteroid's crust and mantle \citep{asphaug2014}. The metallic nature of the asteroid is strongly supported by the value of its high radar albedo, 0.42 $\pm$ 0.1 \citep{shepard2010}, which along with the surface density to be estimated 3,750~kg~m$^{-3}$, suggests 50 per cent macroporosity of the upper 1~m subsoil \citep{ostro1985,magri1999}. 

According to the Bus-DeMeo taxonomy \citep{demeo2009}, the M-types are divided into the X$_{c}$, X$_{k}$ and X$_{e}$ classes of the X--complex and thus Psyche is classified as an X$_{k}$ asteroid in this modern taxonomy. 
Observations in Near-IR have shown features (1--3$\%$ depth) in the pyroxene 0.9~$\si{\um}$ region, having a variation of the band centre and depth at different rotation phases \citep{hardersen2005,sanchez2017}. Moreover the pyroxene absorption band  is persistent with rotation, although the signal is weak, meaning that the silicates that are dominated by orthopyroxene are mixed with the metallic component of Psyche's surface \citep{sanchez2017}. This observation could indicate an endogenous origin of the silicates, but an exogenous origin of these silicates cannot be also ruled out, as we see in the case of Ceres and the presence of enstatite material \citep{vernazza2017}.

A striking interesting result is the detection of the 3~\si{\um} absorption feature \citep{takir2017} on the spectra of the asteroid. This was detected in four different rotation phases, which is consistent with the presence of OH- and or H$_{2}$O-bearing phases. The shape of the 3~\si{\um} feature is similar to other outer Main Belt asteroids and corresponds to the so-called ``sharp-group'' connected to CM-like phyllosilicates \citep{takir2012, rivkin2015}. Moreover, in one dataset the feature is similar to that observed on asteroid Ceres \citep[][Ceres-like group]{takir2012}. This new discovery eventually added Psyche to the list of M-types with 3~$\si{\micro}$m features \citep{merenyi1997,rivkin2000}.  

The presence of hydration sets new constrains on the study of the asteroid. If Psyche is an iron core then any hydration cannot be endogenous but must have been deposited to the surface by another process. Recently it has been demonstrated \citep{mccord2012,turrini2014,turrini2016} that the dark regions on asteroid (4) Vesta are the result of deposition of exogenous material via impacts. It could not be explained as exposition of mantle material after impacts since these events could not be so energetic to excavate in such depth. In addition, the previous model can also verify that the presence of sub-microscopic siderophile elements as exogenous inclusions, originated from hypervelocity impacts on Vesta \citep[see][and references therein]{turrini2014}.

The deposit of exogenous materials on the surface of Ceres can also explain recent observations of this dwarf planet, which were carried out from the Sofia airborne observatory \citep{vernazza2017}.
Contamination of asteroid surfaces from the impact of materials originating from other asteroids is also consistent with recent laboratory findings, where it has been shown that a significant amount of the impactor material can be preserved on the target \citep{daly2015,daly2016, me2016, me2017}. More specifically, it has been shown that the impactor is not pulverised and can be implanted into and around the impact crater at impact velocities $\lesssim$2~km~s$^{-1}$ \citep{me2016, me2017}.  

Here, we study the potential sources of impacts that can deliver OH- and or H$_{2}$O-bearing minerals on Psyche.
The intrinsic impact probability with all possible impact velocities between all possible pairs of asteroids in the Main Belt is 2.86$\times$10$^{-18}$ impacts~km$^{-2}$~yr$^{-1}$ \citep{bottke2005}, resulting to an average impact probability of about 3.6$\times$10$^{-8}$ impacts Myr$^{-1}$ using the most recent estimation for Psyche's size (diameter $D$=225.5 km; see Section~\ref{background}) and sizes of impactors of about 1~km in diameter. As the total impact probability is proportional to the time and the number of available projectiles, this requires a large number of impactors ($\sim$6$\times$10$^{3}$ or more) carrying hydrated minerals and/or water in order to have some deposits on this asteroid over the 4.6~Gyr-long age of the Solar System (which is also likely the age of Psyche).  
While the number of asteroids with spectroscopically measured hydration is growing, their number is very limited: there are 48 known asteroids that display the 3~\si{\um} absorption feature \citep{rivkinphd1997,rivkin2000,rivkin2010,rivkin2015,takir2012} and 28 asteroids are Ch- and Cgh-types \citep[see the database of][at mp3c.oca.eu]{delbo2017}, which are indicative of the presence of the 0.7~\si{\um} absorption band. The 0.7~\si{\um} feature indicates the presence of phyllosilicates likely due to aqueous alteration \citep{vilas1989}.

In order to have a large number of projectiles, our working hypothesis is that these materials were deposited on the surface of Psyche by impacts of asteroids that were members of collisional families whose asteroids have signature of hydrated minerals and/or water-ice.  After the breakup of an asteroid parent body, large number of small fragments -- the family asteroid members -- are produced, thus having an impactor population large enough to overcome the aforementioned low impact probabilities. In addition, it is known that members of asteroid families are broadly compositionally homogeneous, implying that if hydration is observed for one member -- typically the family parent body -- it is reasonable to assume that all the members of the family are hydrated.

As we shall demonstrate, asteroids belonging to the so-called background, namely those that are currently not linked to any known family, also impacted Psyche. There is growing evidence that the background is mostly composed by members of families that are yet to be identified \cite[it has been also predicted by \cite{zappala1996}]{delbo2017, tsirvoulis2017}.
However, at the moment we do not know these ``missing families''. Therefore asteroids of the background, contrary to members of families, do not have a composition link between them and with common parent bodies. 
This makes difficult the estimation of the amount of hydrated material, if any, that impactors from the background population delivered to Psyche.

Our work is structured as follows: in Section~\ref{background} we discuss some relevant physical characteristics of Psyche, such as its mass and size; in Section~\ref{sources} we present the potential asteroid family sources of the impactors; in Section~\ref{model} and Section~\ref{results} we show the method followed and calculate the amount of implanted mass, while Sections~\ref{discussion} and \ref{conclusions} give a discussion and the main conclusion of the results. There, we also show that the so-called asteroid background with low albedo contributes implanting mass on Psyche. However, this background that consists of asteroids currently not associated with any family, cannot be proven that is hydrated, as a genetic relation to the few hydrated bodies in the Main Belt is not yet established.  

\section{Psyche background}
\label{background}
The estimation of the bulk density of Psyche depends obviously on the measurement of the object's dimensions. Several works have been focused on the measurement of the size of Psyche using IRAS observations \citep{tedesco2002}, WISE data \citep{masiero2012}, Akari thermal infrared data \citep{usui2011}, Speckle interferometry \citep{cellino2003}, adaptive-optics imaging \citep{drummond2008}, radar imaging \citep{shepard2008}, and the combination of the available shape model \citep{kaasalainen2002} with occultations \citep{durech2011} and thermophysical models applied to long-baseline, ground-based interferometric observations obtained from the ESO VLTI \citep{matter2013}. 
However all the above measurements lead to densities spanning in a wide range of values from 1,800 $\pm$ 600~kg~m$^{-3}$ \citep{viateau2000} to 6,580 $\pm$ 580~kg~m$^{-3}$ \citep{kuzmanoski2002} implying different composition models and thus are not conclusive. 
A recent work based on radar delay-Doppler, adaptive optics and stellar occultations found the diameter $D_{eff}$~=~226 $\pm$ 23~km, with exact dimensions $279\times232\times189$~km $\pm$ 10 per cent \citep{shepard2017}, which is in an almost perfect agreement with a simultaneous work that constrained Psyche's effective diameter to be $D_{eff}$~=~225 $\pm$ 4~km using also adaptive optics and occultations \citep{hanus2017}. 
Combining this $D_{eff}$ with the most recent mass estimations \citep{carry2012,fienga2014}, the bulk density is calculated to be between 3,500 and 4,500~kg~m$^{-3}$\citep{hanus2017,shepard2017}.
All data combinations of these physical properties, obtained from the latest observations, are given in Table~\ref{densities}.
\begin{table*}
\centering
  \caption{Estimated masses and diameters for Psyche, along with their derived bulk densities. For each value it is given the corresponding reference.(* This value is calculated in this work as an average of Cases 1\&2. **Recent calculations, presented in this work.)}
  \label{densities}
  \begin{tabular}{|c|ccc|}
\hline
\hline
Case & Mass~$\pm1\sigma$  & Diameter~$\pm1\sigma$  & bulk Density~$\pm1\sigma$ \\
&(10$^{19}$ kg) & (km) & (kg~m$^{-3}$) \\
\hline
1. & 2.72 $\pm$ 0.75  \citep{carry2012}& 226 $\pm$ 23 \citep{shepard2017}& 4,530 $\pm$ 1280 \cite{shepard2017}\\
2. & 2.22 $\pm$ 0.36  \citep{fienga2014}& 225 $\pm$ 4 \citep{hanus2017} & 3,700 $\pm$ 630 \citep{hanus2017}\\
3. & 2.54 $\pm$ 0.26  \citep{Viswanathan2017} & 225.5 $\pm$ 4* & 4,200 $\pm$ 490** \\
4. & 2.21 $\pm$ 0.21  & 225.5 $\pm$ 4* & 3,500 $\pm$ 400** \\
\hline
\hline
\end{tabular}
\end{table*}

\subsection{High and low-density scenarios}
Considering as bulk density the value 4,500~kg~m$^{-3}$, which is one of the highest measured in the asteroid population out of those asteroids that have good quality of data \citep[see][]{carry2012}, it strengthens the hypothesis that Psyche could be an exposed metal core of a differentiated asteroid \citep{elkins2017}. According to the models of asteroid differentiation, the process that led to the formation of Psyche happened very early. Considering Psyche's current diameter, $D_{eff}~=~$226~km \citep{shepard2017}, the Psyche parent body (PPB) was supposed to be $\sim$500~km in diameter and have suffered severe ``hit-and-run" impact events capable to remove all crust and mantle, exposing the core \citep{elkins2016}. In addition Psyche should have $\sim$40 per cent macroporosity, if we assume that it is made of blocks of iron/nickel with density around 7,500~kg~m$^{-3}$. In that case the core itself was possibly destroyed and re-accumulated, implying a severe collisional history.  
When an asteroid is catastrophically disrupted, remaining with a mass $\leq$50 per cent of the initial one, after a collision with another body, an asteroid family is formed.
If the collision happened in the Main Belt, a family of asteroid fragments should be in the region of Psyche; however, no family related to Psyche has been found yet \citep{davis1999}. One possibility to solve this issue is that the potential Psyche asteroid family was created at an early time e.g. within the first 500~Myr of the solar system history \citep{davis1999}. This would allow the family fragments to be grounded down by the collisional evolution and be unobservable today. The same models show that, even in this case, there should be today several surviving fragments having diameters around 20~km and being above the detection limit. 
There is a lack of primordial asteroid families in the Main Belt \citep{broz2013,spoto2015}, very likely due to the classical methods that are used to identify them. The Hierarchical Clustering Method (HCM) is not sensitive to find old and dispersed families, as it searches for asteroids forming compact groups in the orbital element space (semi-major axis, eccentricity and inclination). A new approach has been proposed and implemented with success \citep{walsh2013,delbo2017} as it is able to distinguish very old families, having their eccentricities and inclinations dispersed in space. Therefore a possibility of the absence of a Psyche family could be due to searching biases. But this may be an unlike hypothesis because A-type asteroids that could represent mantle material (almost pure olivine) from differentiated bodies, do not extensively exist in the orbital space related to Psyche, but are randomly distributed in the Main Belt \citep{davis1999,demeo2015}.
In order to study further this puzzling small body, NASA is sending a new Discovery Mission to Psyche. The main goal is to get insight if it is a core of a parent body and understand the procedures of differentiation, making all the above questions more valid than ever. The alternative theory is that Psyche is a planetesimal that bares primitive unmelted material \citep{elkins2016}.

However, more recent estimations give lower values for the mass of Psyche. Specifically here we present results for the density of Psyche, using masses that were measured with gravitational perturbations with Mars \citep[construction of the planetary ephemerides INPOP17a, see][]{Viswanathan2017}. The Mars' orbit is significantly perturbed by the asteroids of the Main Belt and lately masses of the most perturbing asteroids have been estimated using Mars orbiters' navigation data \citep{2002A&A...384..322S, Somenzi2010, Folkner2014}. However, such mass determinations are sensitive to the weighting scheme used for the construction of each planetary ephemerides and consequently, one should expect an uncertainty of at least 10 per cent on the mass of Psyche. From such observations densities for this specific asteroid were found to be between 3,500--4,200~kg~m$^{-3}$.

Using the most conservative scenario for the bulk density to be 3,500~kg~m$^{-3}$, if Psyche is the remnant of an Fe/Ni core with a block density of 7,500~kg~m$^{-3}$, the body should still have almost 55 per cent macroporosity.
From older statistics \citep{carry2012} it was shown that asteroids with mass of about 10$^{20}$~kg are solid bodies with almost no marcoporocity. This implies that Psyche with M$\sim$2$\times$10$^{19}$~kg could be at the borderline of this estimation and both scenarios, of a solid or porous body, can be marginally valid. One of the key parameters to start studying the iron core mystery is to consider the presence of OH/H$_{2}$O on its surface. If we assume the iron core scenario then any water should be exogenous and implanted by a number of impacts. 

In this work we are going to consider the Case~3 from Table~\ref{densities}. The $D_{eff}$ is selected as the average of the most recent observational works \citep{shepard2017,hanus2017}, while the mass is one of the latest estimation \citep{Viswanathan2017} that was derived from close encounters with Mars, giving a bulk density of 4,200~kg~m$^{-3}$ and escape velocity $v_{esc}$~=~0.179~km~s$^{-1}$.

\section{Exogenous repositories}
\label{sources}

Following our hypothesis that the exogenous water and/or hydrated material on Psyche come from impactors that are members of asteroid families, we first investigate which of the known families contain members with water and/or hydrated minerals. 
To do so, we retrieve from the literature all asteroids with the 3~\si{\um} band of any type (sharp, round, Ceres- and Europa-type) \citep[e.g.][etc.]{takir2012,rivkin2015}. We then consider only those that are part of a family, using the family list and their asteroid membership of \cite{nesvorny2015}. These are the families of (2)~Pallas, (10)~Hygiea, (24)~Themis, (31)~Euphrosyne, (137)~Meliboea, (153)~Hilda, (363)~Padua (see Table~\ref{asteroids}).
In the following, for each of these families, we present relevant characteristics, we identify and remove interlopers, and study their size frequency distribution (SFD), which we also extrapolate to smaller sizes than the currently observable completeness size in the Main Belt. This is required because catalogued family classification is in general conservative in order to avoid including background objects and to maintain good separation in orbital elements between different families \citep{milani2014}.
\begin{table}
\centering
  \caption{Main belt asteroids with detected 3~\si{\um} band that are also members of families. These families are thus potential sources of hydrated impactors on Psyche according to our working hypothesis. Note that since we find that (110) Lydia is an interloper inside the Padua family (see text), we reject this family as source of hydrated impactors. The asteroids with 3~\si{\um} band that are not members of families are not reported in this table.}
  \label{asteroids}
  \begin{tabular}{|l|l|}
\hline
\hline
Asteroid Family & Hydrated Members \\
\hline
Themis/Beagle & (24) Themis, (90) Antiope, (104) Klymene \\
Hygiea & (10) Hygiea, (52) Europa \\
Euphrosyne & (31) Euphrosyne  \\
Hilda & (153) Hilda  \\
Meliboea &  (511) Davida  \\
Pallas & (2) Pallas \\
Padua  (Rejected) & (110) Lydia \\
\hline
\hline
\end{tabular}
\end{table}

\subsection{Rejection of interlopers and size distributions of families}
\label{familysfd}
As a first step we try to understand if any of the hydrated family members is an interloper to its family. For this reason we check whether the hydrated asteroids are within the Yarkovsky boundaries \citep{vokrouhlicky2006} of their family. Family members form a V-shape in the $a$~vs.~$1/D$ space, where $a$ is the orbital semimajor axis  and $D$ is the asteroid diameter, due to the Yarkovsky non-gravitational forces. The use of this V-shape in the $a$~vs.~$1/D$ space to better constrain the family members has already been proposed and used as method in several works \citep{walsh2013, nesvorny2015,bolin2017, delbo2017}.

\begin{figure}
\includegraphics[width=\linewidth]{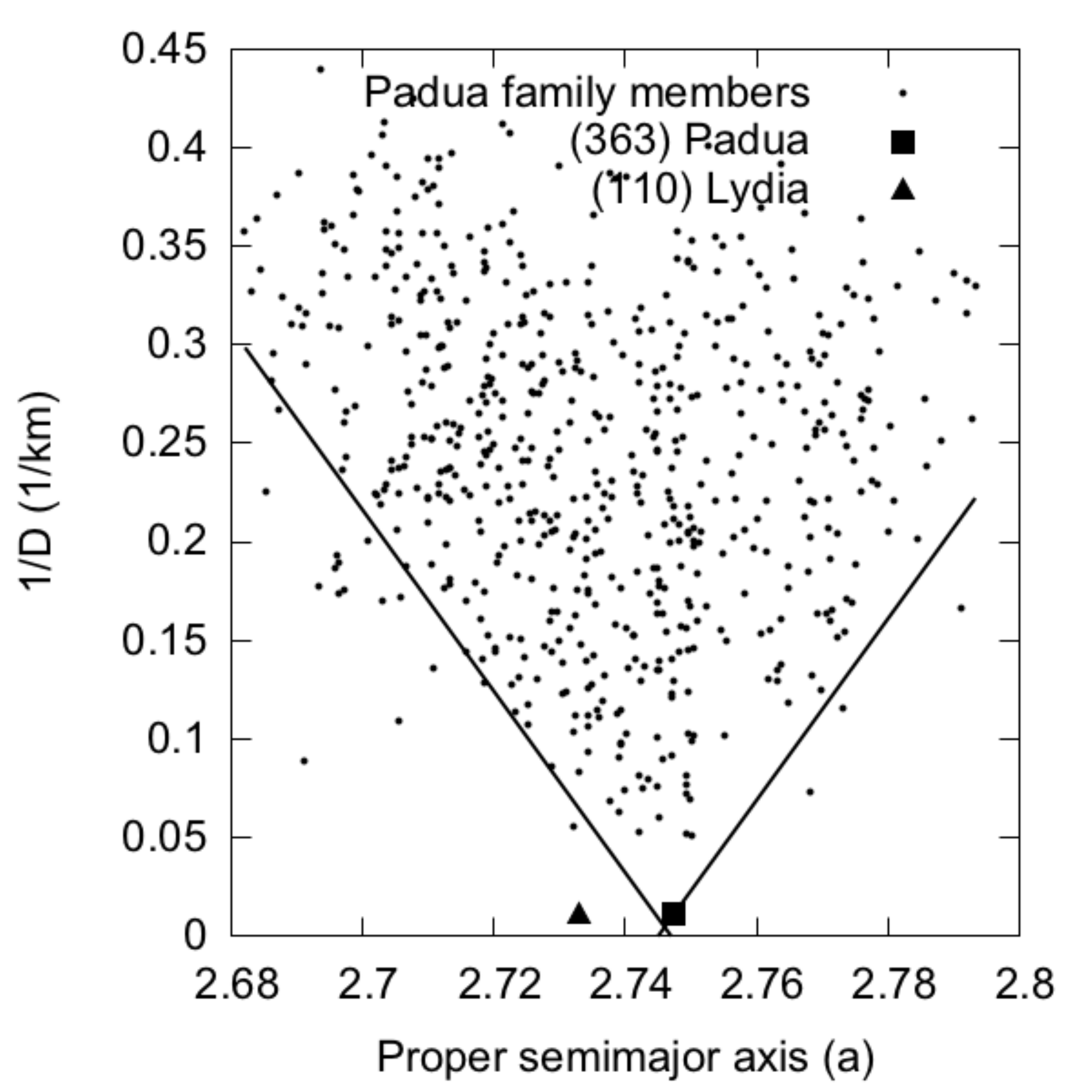}
\caption{Asteroid (110)~Lydia (triangle), which has very similar diameter with (363)~Padua resides out of the V-shape of the family. Along with the spectral and albedo difference, we do not consider it as a member of the Padua family.}
\label{padua}
\end{figure}

In the case of the Padua family, we can clearly identify the hydrated asteroid (110)~Lydia as an interloper. In fact, as it is seen in Fig.~\ref{padua}, Lydia is far out of the inward border of the V-shape. The average albedo of the Padua family is $\left < p_{V} \right >$~=~0.058 and the standard deviation is 0.024. The albedo of (110)~Lydia is 0.17 $\pm$ 0.02 which is more than 3$\sigma$ away from the average albedo of the Padua family.
On the other hand, the albedo of the (363)~Padua is 0.053, matching perfectly the average value. An additional reason that allows us to reject (110)~Lydia from the family is that it has a different spectral classification: (363)~Padua is member of the C-complex, while (110)~Lydia is classified as an M-type (or X$_{k}$-type). Since no other Padua family member has a detected 3~$\si{\micro}$m band, we exclude this family as a potential source of hydrated impactors on the surface of Psyche.

The cumulative size frequency distributions (SFDs) are obtained using the same database of asteroid physical properties of \cite{delbo2017}. If the diameter is not reported in the literature we calculate it from $D$(km)~=~$(1329/ \sqrt{\left<p_V\right>})\times 10^{-H/5}$ \citep[see][and references therein]{harris1998} where $\left<p_V \right>$ is the average albedo of the family members and $H$ the absolute magnitude of each object.
\begin{figure}
\includegraphics[width=\columnwidth]{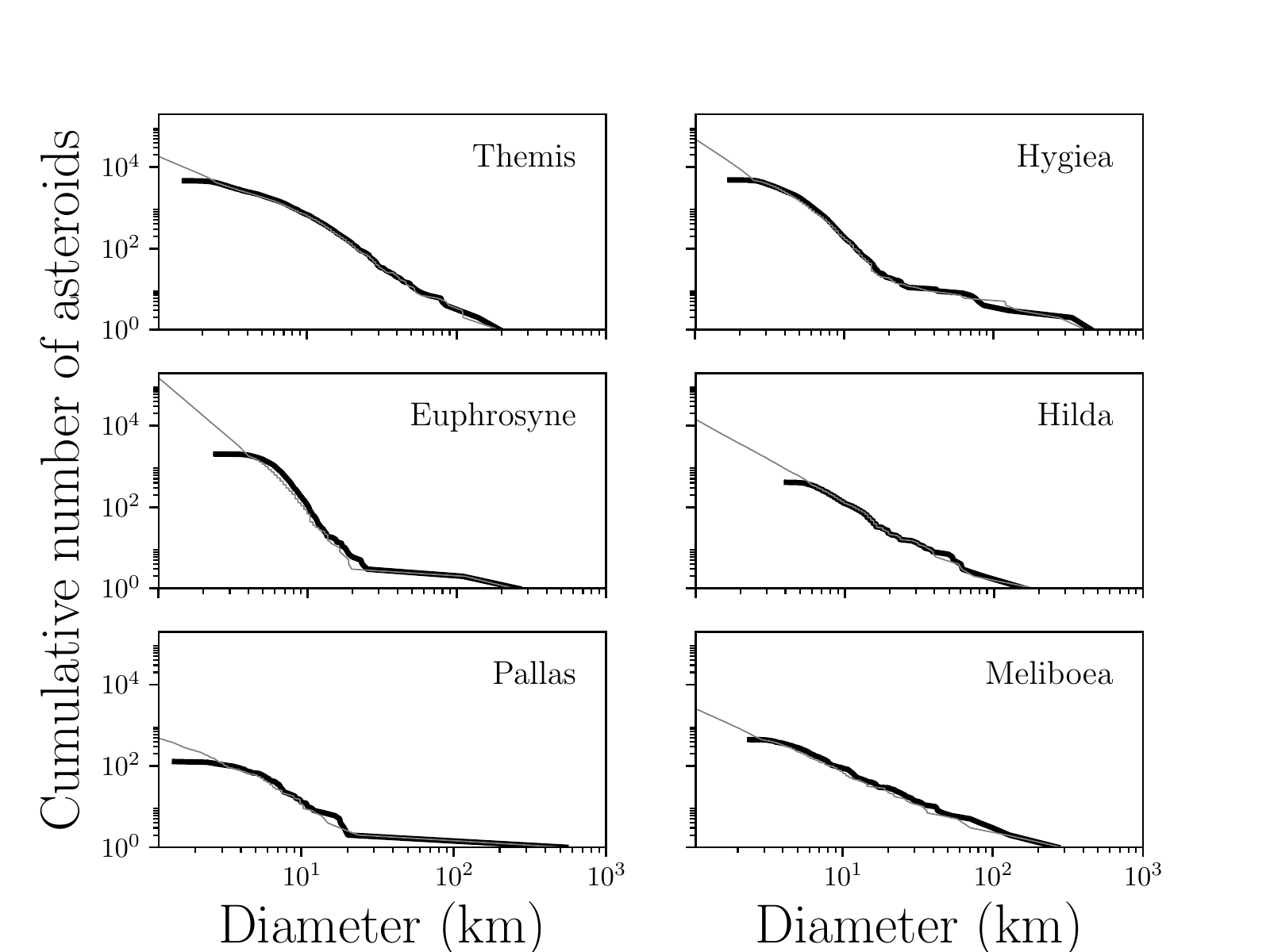}
\caption{The observed SFDs (black line) of the six families containing members with detected 3~\si{\um} band. Their slope (grey line) is calculated by fitting a power law function of the diameter to the observed population with $D\geq D_{17}$, where $D_{17}$ is the diameter that corresponds to $H$~=~17, the current completeness limit of the Main Belt.}
\label{sfds}
\end{figure}

The observed populations of the families, however, are not complete due to observational biases, as smaller asteroids become more difficult to be detected by telescopic surveys. The current limit of completeness of catalogued main belt asteroids is $H \sim$~17 \citep{jedicke2015,denneau2015}. Therefore we consider the SFD of the families as complete for $H<$~17.  This $H$ limit corresponds to the diameter limit $D_{17}$, which value depends on the average albedo of the family (see Table~\ref{slopes}). 
We assume that the cumulative number of asteroids down to any diameter $D<D_{17}$ is given by:
\begin{equation}
N = N_{17} D^a {D_{17}}^{-a}
\label{eq1}
\end{equation}
where $D_{17}$ the diameter at $H=17$ and $N_{17}$ the total number of bodies with $H<$17.
The slope $\alpha$ is obtained by fitting a power law of $N(D)$~=~$k D^{\alpha}$ (which is a straight line in a $\log N$ vs. $\log D$ space) to the known population for $H<$~17 and is negative as the d$N$/d$D$ increases for decreasing $D$.

\begin{table}
\centering
  \caption{The slope $\alpha$ of each family is calculated by fitting a power law to their observed SFD up to the completeness ($D_{17}$) which actual value depends on the average albedo of the family members.}
  \label{slopes}
  \begin{tabular}{|l|l|l|l|l|}
\hline
\hline
Asteroid Family &$ <p_{V}>$ & $N_{17}$ & $\alpha$ &$D_{17}$(km)\\
\hline
Themis & 0.069 & 4535 &-1.52& 2.00\\
Hygiea & 0.072 & 4837 & -2.30 & 1.97\\
Euphrosyne & 0.060 & 2035 &-2.91& 2.21  \\
Hilda & 0.051 & 409  &-1.99& 2.34\\
Meliboea & 0.066 & 444 &-1.56& 2.06\\
Pallas & 0.142  & 128 & -1.32 & 1.40\\
\hline
\hline
\end{tabular}
\end{table}

\subsection{Themis and Hygiea families}

Themis and Hygiea asteroid families are considered as the major contributors of impactors on Psyche because they have the most numerous family members and the 3~\si{\um} band is detected not only on the parent body but on several objects per family.

Themis is a family of the outer Main Belt ranging from 3.02 and 3.24~au in proper semimajor axis ($a_p$), between 0.0116 and 0.0404 in proper sin inclination ($sin(i_p)$) , and between 0.1115 and 0.1862 in proper eccentricity ($e_p$). It was formed 2.5~$\pm$~1.0~Gya \citep{broz2013}, however, a recent estimate gives the age limits for Themis to be between 2.4 and 3.8~Gyr \citep{spoto2015}. The majority of the Themis family members belong to the C-complex \citep{florczak1999,deleon2012,fornasier2016}. This family is located in an area where hydration \citep{florczak1999,takir2012} and water ice have been found. The observation of the 3~$\si{\micro}$m band on (24)~Themis, the family parent body, could be explained by a 4 per cent of water-ice on the surface \citep{rivkin2010} that is additionally distributed in a wide range of latitudes \citep{campins2010}, made it the first C-type asteroid having a water-ice layer with organic compounds on its surface.
The presence of water-ice in the Themis family is consistent with the discovery of the so-called Main Belt Comets (MBC) \citep{haghighipour2009,hsieh2012,hsieh2014} within this family. The most probable explanation for the activity of the MBC is that the later is driven by sublimation of water-ice and other volatiles.
Apart from the MBCs other asteroid members of the family were found to have both the ``round" 3~\si{\um} water-ice band, such as (90)~Antiope \citep{hargrove2015} and the ``sharp" absorption such as (104)~Klymene \citep{takir2012}, showing the presence of OH into the mineral lattice \citep{jones1988}. Additional evidence, that these objects could have high water content, comes from the densities of three Themis members, (24)~Themis, (90)~Antiope and (379)~Huenna, that have very low density values, around 1,280--1,800~kg~m$^{3}$ \citep{descamps2007,carry2012}. 
Geophysical evolution models for (24)~Themis \citep{castillo-rogez2010} have suggested that the parent body had accreted from ice mixed with carbonaceous material and before the time of the breakup -- forming the Themis family -- it differentiated, implying that more asteroids members of the family should contain water-ice. Recent compositional characterisation of Themis family members confirms the above scenario. Assuming a parent body of 400~km in diameter its accretion should have happened about 4~Myr after the CAIs and the aqueous alteration happened in the core and not in the outer shell. Thus a number of Themis members should consist of ice and anhydrous silicates while others of hydrated minerals \citep{marsset2016}. 
This scenario is consistent with a previous study that showed that two thirds of the C-complex asteroids of the outer Main Belt should have undergone aqueous alteration \citep{jones1990}.
The mean geometric visible albedo of the family is $\left < p_V \right >$~=~0.069 with an rms of 0.02. According to the family classification of \citep{nesvorny2015} it has a total of 4,633 members.

The family of Hygiea is younger than Themis and was formed after a cratering collisional event 2.0~$\pm$~1~Ga \citep{broz2013} (the family age of $\sim$1.3~$\pm$~0.3~Gyr from \cite{spoto2015} agrees with \cite{broz2013} estimates). This family is also located in the outer Main Belt (3.029~$<a_p<$~3.242~au), where inclinations of its members (0.0691~$<sin(i_p)<$~0.1167) are larger than those of Themis family members, but the eccentricities of the two families (0.0941~$<e_p<$~0.1722) almost overlap. On (10)~Hygiea itself and (52)~Europa the absorption band has been detected. However, according to the current classification \citep{takir2012}, (10) Hygiea has a band typical of the Ceres-group while (52) Europa has a different band, constituting the ``Europa-group''.
The mean geometric visible albedo $\left < p_V \right >$ of the family is 0.07 with an rms of 0.03, and it has a total of 4,852 members \citep{nesvorny2015}.


\section{Collisional and contamination model}
\label{model}
Having identified the asteroid families that are the potential sources of impactors with a 3~\si{\um} band, in this section, we describe the method to calculate the amount of the implanted material on asteroid Psyche, from each of these sources. We broadly follow the technique that has been developed to explain the dark deposits on the surface of the asteroid (4) Vesta \citep{turrini2014}, that includes a collisional and contamination model. 

The number of impact events on Psyche for a specific diameter bin of width d$D$ of the impactor population can be calculated as \citep{2011obrien}:
\begin{equation}
d N_I = P_{P}~A~\frac{dN}{dD}~T dD,    
\label{eq2}
\end{equation}
where $P_{P}$ the intrinsic impact probability (per year and per km$^2$) on Psyche, $A$ the cross-section of the target and of the impactor of size $D$, d$N$/d$D$ the differential size distribution of the impactor population and $T$ the time interval. The cross-section $A$ is calculated as:
\begin{equation}
A = (R_{P}+0.5D)^{2} \sim R_P^2,
\label{eq3}
\end{equation}
where $D$ is the diameter of an impactor and $R_{P}$~=~112.5~km the current estimation for the radius of Psyche (Table~\ref{densities}). 

The intrinsic impact probability of each source of impactors on Psyche, $P_{P}$, depends on the orbits of the target and of the impactors. We calculate $P_P$ using the algorithm of \cite{wetherill1967}, which, given the semimajor axis, eccentricity and inclination of the orbits of two asteroids (a target and an impactor) and assuming that mean anomaly, argument of perihelion and longitude of the ascending node have a uniform probability distribution over the interval 0~--~2$\pi$, computes which fraction of these angles corresponds to the two objects being closer to each other than 1~km. Next, the algorithm translates this fraction into an intrinsic collision probability per km$^2$ and per year ($P_{P}$) using the orbital periods of the two objects and assuming that they are not in resonance with each other. For each possible impact configuration (e.g. when target and projectile are closer than 1~km), the algorithm also calculates the mutual (impact) velocity $V$. In a final step, the probabilities of all possible impact configurations are summed and the code determines an average impact velocity. 
We apply this algorithm to estimate the value of $P_{P}$ and $V$ between Psyche and each family member. For each family, we average the values of the $P_P$ and of $V$ among its members (Table~\ref{prob}).

For $T$, we use the estimated ages of the families taken from the literature. In some cases the age is not well determined, thus we perform the estimation of the number of impacts on Psyche and the amount of implanted material, for each age case (see Table~\ref{masses}). In particular, we consider an upper and a lower limit for the ages of the Themis, Hygiea and Euphrosyne families. We follow a scenario in which a family contributes to the impactors flux on Psyche throughout the family lifetime $T$ from Eq.~\ref{eq2}, which is equal to the family age.

The implanted mass $m$ is a fraction $f$ of the impacting mass $M$ on Psyche \citep{svetsov2011}. The value of $f$ is given by: 
\begin{equation}
f = (0.14 + 0.003V)\ln v_{esc} + 0.9 V^{-0.24},
\label{eq4}
\end{equation}
where $v_{esc}$~=~0.179 km~s$^{-1}$ the escape velocity from Psyche. For the value of the impact velocity $V$ we take the average impact velocity of each family obtained in the previous step and is given in Table~\ref{prob}. 

\begin{table}
\centering
  \caption{Average impact probabilities and collisional velocities of the source families on Psyche. The fraction of the retained mass depends upon the impact velocity $V$.}
  \label{prob}
  \begin{tabular}{|l|l|l|l|}
\hline
\hline
Asteroid  & $<P_{P}> \pm \sigma$ & $<V> \pm \sigma$ & $f$\\
Family & (yr$^{-1}$km$^{-2}$) & (km~s$^{-1}) $ & \\
\hline
Themis & 7.9 $\pm$ 1.0$\times$10$^{-18}$ & 3.0 $\pm$ 0.1 & 0.41\\
Hygiea & 4.1 $\pm$ 0.53$\times$10$^{-18}$ & 3.1 $\pm$ 0.2 & 0.40\\
Euphrosyne & 2.19 $\pm$ 0.53$\times$10$^{-18}$ & 8.8 $\pm$ 0.3 &  0.22 \\
Hilda & 5.8 $\pm$ 5.8 $\times$10$^{-19}$ & 2.0 $\pm$ 2.0 & 0.49\\
Meliboea & 2.5 $\pm$ 0.44 $\times$10$^{-18}$ & 5.6 $\pm$ 0.2 & 0.30\\
Pallas & 1.7 $\pm$ 0.13 $\times$10$^{-18}$ &  10.7  $\pm$ 0.2 & 0.19\\
\hline
\hline
\end{tabular}
\end{table}

The number of impacts onto the target is governed by Poisson statistics, and therefore is affected by an uncertainty of the order of the square root of the number of the calculated impacts. In order to take this into account we calculate, for each asteroid family, the number of objects that are needed in order to give five or more impacts on Psyche, during their lifetime. This translates to 99.9 per cent probability to have impact events and is calculated from the cumulative probability to obtain one or more events from a Poisson distribution with an average success rate of five. 
To achieve this flux of impacts, the required number of projectiles $N_{lim}$ is given by $5 = P_{P}~A~T~N_{lim}$. As the number of objects (and of impacts) increases with decreasing size, given the size distributions of asteroids (Fig.~\ref{sfds}), we determine for each family the value of $D_{lim}$, which is the diameter value at which the cumulative SFD of each family has a number of asteroids equal to $N_{lim}$. 
This value of $D_{lim}$, obviously, depends on the age of the family -- younger age requires larger number of projectiles in order to achieve the 5 impacts and thus is needed to extrapolate to smaller diameters (see Table~\ref{masses}). 

The implanted mass from each family on Psyche is calculated in two steps, using a Monte Carlo procedure for the first part of the SFD for $D>D_{lim}$ and an analytical method for $D<D_{lim}$. For the first part of the SFD (where $D$>$D_{lim}$) we also set an upper limit to the size of the population to be $D^*$. The value of $D^*$ represents the smallest impacting body that could catastrophically disrupt our target Psyche \citep{bottke2005}. For each family used here this limit is slightly different as it depends on the average impact velocity that the members of each family have on Psyche (Table~\ref{prob}), as detailed before. We obtain the value of $D^*$ from Fig.~13 of \citep{bottke2005}, by interpolating the value of $Q^*$, the intrinsic energy required to disrupt an asteroid, across the curves plotted for different impact velocities and for a target having the diameter of Psyche (Table~\ref{masses}). 
 
\begin{figure}
\includegraphics[width=\columnwidth]{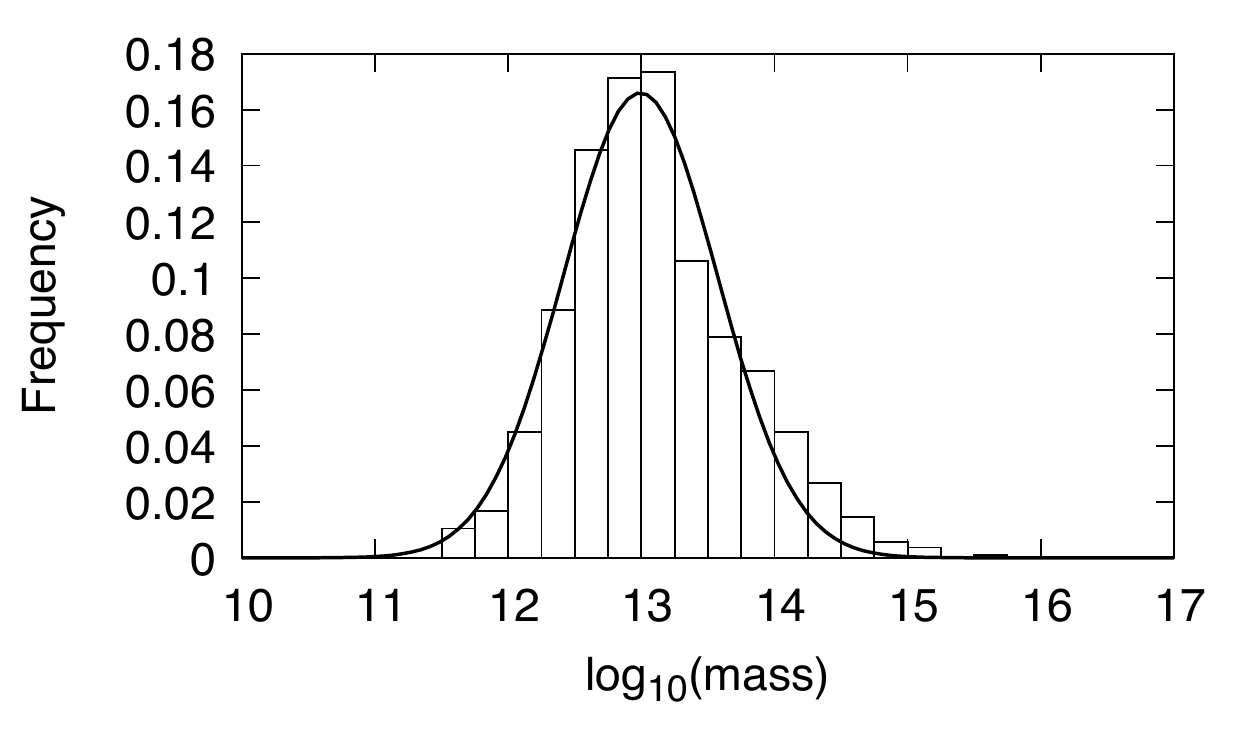}
\caption{The result of 10$^{4}$ Monte Carlo simulations for Hygiea family for $T$=2~Gyr. The average implanted mass is 9.7$\times10^{12}$~kg.}
\label{distr}
\end{figure}

For each family, the implanted mass from the size distribution between $D_{lim}$ and $D^*$, is obtained using a Monte Carlo method. Firstly we extrapolate the observed size distribution from the completeness limit $D_{17}$ to $D_{lim}$ by generating a random number of synthetic bodies with a size distribution, tied to the observed one, and with a slope $\alpha$ as given in Table~\ref{slopes}. Next, we take a Monte Carlo step of 100~Myr, during which a random number is generated uniformly between 0 and 1 for each family member of the observed and extrapolated populations. If this number is \ca{smaller} than  $P_P \times A \times 100$~Myr, then the asteroid is considered to have impacted Psyche and is removed from the simulations. In case of an impact, we store the value of the impactor's diameter. We perform a number of steps to reach the age of the family. Once the Monte Carlo simulation is complete, we calculate the implanted mass by: 
\begin{equation}
m_{D_{lim},D^*} = f ~ \rho \frac{ \pi }{6} \sum_{i=0}D_i^3,
\label{eq6}
\end{equation}
where $i$ runs over the bodies that were marked as impacting.
Since this is a stochastic process, involving relatively low number of impactors, we repeat for each family, each Monte Carlo run 10$^{4}$ times in order to estimate the average implanted mass and its deviation (Fig.~\ref{distr}). We find (Section \ref{results}) that the distributions of the implanted mass are quasi-log-normal. Therefore, we fit a Gaussian function to the occurrence distribution of the $\log_{10}$ of the implanted mass, allowing us to derive the most probable value and its standard deviation, which are presented in Table~\ref{masses}. 

For the rest of the SFD, for each source population of impactors, the d$N$/d$D$ is given by the size distribution of each family as described in Section~\ref{familysfd}. We write that:
\begin{equation}
\frac{dm}{dD} = \frac{dN}{dD}~f T A P_P~\rho \frac{ \pi  }{6} D^{3},
\label{eq5}
\end{equation}
which results from d$M$/d$D$~=~d$N$/d$D$$\times$d$M$/d$N$ and from d$M$/d$N$~=~$\rho \frac{ \pi  }{6} D^{3}$.

From the dust particles with $D=0$ to $D=D_{lim}$, the implanted mass is given by:
\begin{equation}
m_{0,D_{lim}}  = f T A P_P ~ \rho \frac{ \pi  }{6} ~ \alpha D_{lim}^{-\alpha} N_{lim} \int_0^{D_{17}} D^{(\alpha + 2)} dD.
\label{eq7}
\end{equation}
This function is integrable for $\alpha > $~-3 and results in the solution:
\begin{equation}
m_{0,D_{lim}} = - f T A P_P~\rho \frac{ \pi  }{6} ~ \frac{\alpha D_{lim}^{-\alpha} N_{lim}}{\alpha +3 } D_{lim}^{(\alpha + 3)}.
\label{eq8}
\end{equation}

To convert sizes to masses we assume spherical asteroids and constant bulk density value for all the family members. We collect from literature \citep{hanus2017, fienga2014, carry2012} densities of asteroids that belong to the families that mainly contribute to collisions (Themis and Hygiea) and average them. More specifically, using data for (10)~Hygiea, (24)~Themis, (90)~Antiope, (52)~Europa and (379)~Huenna we obtain an average density of $\rho$~=~1,520~kg~m$^{-3}$. This value is also in agreement with estimates of C-complex asteroids densities from \cite{delbo2017}.
The total implanted mass is $m_{0,D_{lim}} + m_{D_{lim},D^*} $ summed over all the asteroid families that contribute to impacts on Psyche. 
In our model, older the family more time this source region had to implant mass on Psyche, hence higher the implanted mass. Since for some of the hydrated families the age is very uncertain, we calculated the implanted mass for an upper and lower limit of their ages, when multiple estimations exist.

\section{Results}
\label{results}

We start showing, with very simple arguments, that there is high probability that asteroids from the Themis family alone have impacted Psyche, thus delivering hydrated/icy material. This holds when we conservatively restrict ourselves to the number of asteroids presently linked to the Themis family, using the family identification of \cite{nesvorny2015}. Table~\ref{prob} shows that the total probability that an asteroid of the Themis family has to impact Psyche is 7.9$\times$10$^{-18}$~yr$^{-1}$~km$^{-2}$, resulting in 1.8 impacts when the number $N$~=~4,636 of known family numbers is multiplied by $P_P$, by the cross section of Psyche, and by the upper limit of the age of Themis, $T$~=~3.8~Gyr. This estimation of the number of impact events is governed by Poisson statistics, and therefore the cumulative Poisson probability to have 1 or more impacts is 0.8347, given the 1.8 average value for impacts. The number of impacts can be even higher (2.2) when the number of family members identified by \cite{milani2014} (AstDys download of Nov 2017, $N$~=~5,612) is used, resulting in an impact probability of 0.8892.

Following the method described in Section~\ref{model}, we calculate the implanted masses for each hydrated asteroid family. Table~\ref{masses} gives our results for each part of the SFD as described in the previous section. The total averaged mass that is implanted, summed over all the source asteroid families and calculated using the maximum and minimum family ages, is in the range $m_{total}$$\simeq$6.2--13.6$\times$10$^{13}$~kg, meaning that when we consider the upper age limits for Themis, Hygiea and Euphrosyne the implanted mass is more than double. The contribution from the Themis family is between $\sim$40--60 per cent of the total amount, depending on its age, and that from Hygiea is $\sim$7--10 per cent. In addition, we see that the Euphrosyne family contributes a large amount of mass, $\sim$29 per cent (for both age estimates), primarily from the small-size impactors where $D<D_{lim}$, which can be explained due to its steep SFD (it gives the largest number \ca{of} bodies when extrapolated to small sizes). However, when we consider $D>D_{lim}$, its lower $P_P$ and the high value for the impact speed lead to less implanted mass on the target, which is just a small fraction of the one coming from Themis and Hygiea. In this calculation we assume that the slope of Euphrosyne is constant to dust ($D=0$), but this may not be the reality. It should be also mentioned that  according to Eq.~\ref{eq8} the mass of a family is finite for slopes greater than --3 and Euphrosyne is very close to this limit.

The Pallas, Meliboea and Hilda families contribute very little, about 1 per cent, compared to Themis and Hygiea, which are the main impactor sources on Psyche. This can be explained in several ways. First of all the later two families are the most numerous, with the higher impact probabilities and lowest impact speeds. As it is shown, the mass is affected by the collisional velocity (see Eq.~\ref{eq4}) and in our case Themis provides the flux of impactors with the lowest one. 

\begin{table*}
\centering
\caption{Family ages from the literature (Age) and adopted ($T$) in our model to calculate the number of impact events on the asteroid Psyche. For Hygiea, Themis, and Euphrosyne we run our model considering two possible $T$ values for their age, high (H) and low (L). The implanted masses are calculated for the two parts of their SFDs (see text for further details). The mass range from the families of Themis, Hygiea and Euphrosyne is calculated for a maximum (H) and minimum family age (L).}
\label{masses}
\begin{tabular}{|l|ccc|ccc|c|}
\hline
\hline
Asteroid Family & Age & Reference  & $T$ &  $D_{lim}$ & $D^{*}$ & Implanted Mass & Implanted Mass\\
impactor source & & &  &  & & $D$>$D_{lim}$   & $D$$\leq$$D_{lim}$ \\
\hline
\rule{0pt}{3ex}
&(Gyr)  & & (Gyr) & (km) & (km) & $\times 10^{13}$ (kg) & $\times 10^{13}$ (kg)\\
\hline
\rule{0pt}{3ex}
Themis (H) 	& 2.45--3.8 $\pm$ 0.9  & \cite{spoto2015} & 3.8 &  1.30 & 70 & $ 7.84^{+43.60}_{-6.64}$ & 0.37\\
\rule{0pt}{3ex}
Themis (L) 	&  2.5 $\pm$ 1 & \cite{broz2013} & 2.5 & 0.94 & 70 & $ 2.49^{+12.60}_{-2.08} $ & 0.16\\
\rule{0pt}{3ex}
Hygiea (H)  	 & 2.0 $\pm$ 1 & \cite{broz2013} & 2.0 & 1.0 & 71 & $ 0.97^{+2.65}_{-0.71} $ & 0.44\\
\rule{0pt}{3ex}
Hygiea (L)   	& 1.3 $\pm$ 1 & \cite{spoto2015} & 1.3 & 0.82 & 71 & $ 0.47^{+1.26}_{-0.34}$ & 0.24\\
\rule{0pt}{3ex}
Euphrosyne (H)  &  $<$ 1.5 & \cite{broz2013} & 1.5 & 1.06 & 67 & $ 0.40^{+0.63}_{-0.24}$ & 3.55\\
\rule{0pt}{3ex}
Euphrosyne (L)  & 1.1--1.3 $\pm$ 0.3  & \cite{spoto2015}  &  1.1 & 0.95 & 67 &$ 0.27^{+0.41}_{-0.16}$ & 2.58\\
\hline
\rule{0pt}{4ex}
& & &&&&$\times 10^{10}$ (kg) & $\times 10^{10}$ (kg)\\
\hline
\rule{0pt}{3ex}
Hilda 		& $\sim$ 4.0 & \cite{broz2011} & 4.0 &  0.26 & 74 & $ 24.5^{+73.77}_{-18.45}$ & 7.08\\
\rule{0pt}{3ex}
Meliboea       	& $<$ 3.0 & \cite{broz2013}  & 3.0 & 0.15 & 75 & $ 5.01^{+22.0}_{-4.0} $& 0.44\\
\hline
\rule{0pt}{4ex}
& & &&&& $\times 10^{6}$ (kg)  & $\times 10^{6}$ (kg)\\
\hline
Pallas            	& $<$ 0.5 & \cite{broz2013} & 0.5 & 0.005  & 70 & $ 2.23^{+14.68}_{-1.94} $ & 0.074\\
\hline
\hline
\end{tabular}
\end{table*}

This result is probably a conservative scenario, as the used slope for the extrapolation is the one derived from the observed population and not the initial that each family had when was created. This means that, especially for the old families, the slope should have been steeper than the current one and thus we should lack objects from the distribution. Collisional and dynamical depletion of asteroids during the last 3.8~Gyr, have removed probably about 50 (or more) per cent of the Themis initial population \citep{minton2010,delbo2017}.
Any difference in size distribution will lead to difference on the number of impacts and the final amount of the implanted mass. Hence, ours is likely a lower limit of the implanted mass. Another fact that needs consideration is that the Themis and Hygiea families, apart from the main source of impactors, are also the contributors with the lowest impact velocities (Table~\ref{prob}). Therefore any Themis and Hygiea impactor should implant higher amount of mass compared to the rest source regions.

We apply the same technique -- with some differences explained below -- to the background population of asteroids with low albedo, i.e. those bodies currently unlinked to any known family. We consider only those background asteroids of the Main Belt and the Hilda population ($2.1 < a < 3.5$~au) with geometric visible albedo $p_V<0.12$ -- this threshold has been shown to include most of the asteroids of the spectroscopic C-complex -- and with $H<17$. The latter is to avoid including asteroids below the completeness limit of the Main Belt.
This results in 46,126 asteroids and the best fit power law to the cumulative SFD has exponent $-1.81$. The observed SFD of the background asteroids starts to deviate significantly from this power law function at $D$=3.3~km, with the observed SFD rolling over and becoming flat, which indicates that below this size, the sample of asteroids with measured $p_V<0.12$ becomes observationally incomplete. Indeed the completeness limit of the sample of asteroids with measured size and albedo is 2, 3 and 5~km for the inner, middle, and outer part of the Main Belt respectively (Masiero, J., private communication). The point where the best fit power law becomes bigger than the measured SFD is at $D=3.3$~km and $N=38,141$ asteroids.
Assuming that these asteroids have an average intrinsic probability of impacting Psyche of 2.8$\times$10$^{-18}$ impacts per year per km$^{2}$ and an average impact velocity of $v=5.3$~km~s$^{-1}$ \citep[the main belt average value, see][]{bottke1994} we find that the average number of impacts over the age of the Solar System (4.5~Gyr) is 6 with a standard deviation of 2.4. Our 10$^{4}$ Monte Carlo simulations resulted in $6.06 \pm 2.47$ impacts, a result that agrees perfectly with the Poisson statistics, which gives 6.08 impacts with a square root of 2.47.

The corresponding implanted mass, which results from our 10$^{4}$ Monte Carlo simulations, has a probability distribution which core is log-normal, with the high-mass tail significantly above a Gauss distribution. The best fit Gaussian function in log$_{10}$(mass)~vs.~log$_{10}$(probability) space is centred at log$_{10}$(mass) = 14.6 with a standard deviation of 0.4. This implies that the most likely implanted mass by the background is about 4$\times$10$^{14}$~kg, with the most likely range between 1.6$\times$10$^{14}$~kg and 10$\times$10$^{14}$~kg. 
The integral of the implanted mass from the background population that is smaller than $D=3.3$~km, assuming the power law SFD with exponent $-1.81$ and tied to the observed SFD is estimated using Eq.~\ref{eq8}. This results in 8.9$\times$10$^{13}$~kg which is about 1/5 of the mass implanted by the observed population of the low albedo asteroid background.

\section{Discussion}
\label{discussion}
The understanding from the previous calculations is that Psyche has been the target of impactors originated from hydrated sources.
The first outcome is that about 52--68 per cent (considering all age estimates) of the implanted mass is coming from the impactors with $D>D_{lim}$ that have 99.9 per cent probability to produce an event. The contribution of the sub-km objects down to dust sizes corresponds to the rest 32--47 per cent. 

Here we need to comment that it is known that families are in general bigger than those of catalogues. However, due to the fact that the focus of their study was to obtain a good classification of families, the authors of these works adopted a conservative approach in the selection of their Quasi-Random Level (QRL) for the hierarchical clustering analysis \cite{zappala1990}, in order to avoid background objects from being incorrectly identified as family members and maintain good separation in orbital elements between families.

One main question that naturally comes is if the hydration of impactors could survive an impact of speeds representative of the Main Belt (e.g. $\left<V\right>$~=~3~km~s$^{-1}$ for Themis family impactors) and be successfully implanted on the surface of the target.  
Hydrocode simulations using different configurations for the impacting materials, such as bulk density and composition (stony/Fe, Fe/Ni) and low impact speeds (2~km~s$^{-1}$), show that the temperature never exceeds the 150~K at the extreme cases when the impacting body is pure water-ice (Mark Price, private communication).  
\cite{turrini2014life} studied and confirmed with simulations the delivery and survival of hydrated material on Vesta by the impacting asteroid population, following several scenarios for their flux and size frequency distribution. They showed that there is an effective contamination of the target asteroid with small impactors (1--2~km in diameter). Following this work, \cite{svetsov2015} investigated the water delivery to the Moon by asteroids and comets, at higher impacts speeds than the average one in the Main Belt \citep[5.3~km~s$^{-1}$, see][]{bottke1994}. It was shown specifically that the delivery not only can happen but the hydrated components can survive the estimated impact temperatures. Supportive to this scenario is the fact that the temperature inside cm-size impactor's ejecta fragments cannot exceed values responsible for de-volatilization, because the exposure to extreme impact conditions is very short.
Since phyllosilicates have been found in the IDPs (Interplanetary Dust Particles) this means that hydration can survive impact events, followed by ejection and atmospheric entry \citep{rivkin2002}. This implies that hydration can also survive from collisional events that formed the asteroid families \citep[e.g. the Ch and Cgh asteroids in Dora and Chloris families,][]{busphd1999}. 
An additional fact is that it has been shown that the 3~$\si{\um}$ band on Murchison and other carbonaceous meteorites can be removed by exposure to higher temperatures, e.g. $\sim$400--600~$^{\circ}$C \citep{hiroi1996}. So, although an impact could partially remove ice, it is very likely that the hydrated minerals would survive collisions at such speeds.

Another aspect that needs consideration is how much of the implanted mass, coming from hydrated families, on Psyche (10$^{14}$~kg), is actually hydrated and contributes to the 3~$\si{\um}$. From literature we find that the hydration on meteorites can be up to 2--3 per cent \citep[for CI, CM, CO, CR meteorite samples, see][and references therein]{rivkin2000}. In our case this corresponds to 10$^{12}$~kg of hydrated implanted material.
However, the values of weight percentages on asteroid surfaces are lower by an order of magnitude \citep[e.g. the values of the hydrated M-type asteroids in][and references therein]{rivkin2000}, an effect that could be explained by space weathering and loss of surface hydration. 
Can this amount excuse the observations or more impactors are needed and thus more sources should be identified? A supportive factor is that the 3~$\si{\um}$ band on Psyche reflects two different types of hydration \citep{takir2017}, that although theoretically would require aqueous alteration processes at different times or depths on the body, could also indicate different impactors. Here we show that it can be caused by the different types of the impactors. Another possibility is that a part of the hydration of Psyche could occur in the primordial belt, which may have been denser populated and with a population of bodies having a different orbital dispersion. 

Any impact event, besides the implantation of new material, causes erosion of the target. The ratio of erosion/implantation is affected by the impact speed. At speeds above $\sim$2~km~s$^{-1}$ the erosion is always bigger than the retention of the exogenous mass \citep{turrini2014life}. Considering also the fact that higher impact speeds lead to higher ejecta speeds, it is probable to exceed the escape velocity limit and loose material in space. Therefore impacts throughout the history of Psyche implant, bury and remove any previously delivered hydration.

Psyche used to be the only large M-type asteroid that was not hydrated, being an outlier in this spectral group. Since the hydration was confirmed, the boundary that was observed around 60--70~km between the hydrated and non-hydrated asteroids \citep[Fig.~11 from][]{rivkin2000} makes even more sense than before. One explanation would be the effect of the small cross-section $A$ of the asteroids when is applied to the Eq.~\ref{eq2}, as large asteroids can suffer more impacts than the small ones. By checking one by one the diameters of all the 48 asteroids with the 3~$\si{\um}$ band -- which belong to different spectroscopic complexes -- we clearly see that that boundary can be confirmed, with only a few exceptions (e.g. asteroids (261)~Prymno and (418)~Alemannia).

\section{Conclusions}
\label{conclusions}

We study the possibility that the 3~$\si{\micro}$m band of the main belt asteroid (16) Psyche has been caused by the contamination from exogenous material.  We investigate the collisional environment of the asteroid to understand which could be the hydrated sources. Themis and Hygiea asteroid families are good potential candidates, since the hydration band has been observed to several of their members. We find that these families could contribute in total with up to 10$^{14}$~kg of foreign mass, since their formation event. \ca{Investigating the potential contribution from the background asteroid population we found that matching amount of mass ($\sim$10$^{14}$~kg) could originate from that source. However, this is not a definite conclusion as the hydration of the background is not proven.} The total implanted mass is related to the impactor's size and impact speed, and the actual hydrated fraction of it is yet not accurately known. Further laboratory experiments are needed to estimate the hydration required in order to reproduce the observed band. The contamination due to impacts is a working hypothesis which does not exclude both scenarios for Psyche, to be either an Fe/Ni core of a differentiated parent body or a stony/Fe body with high metal component.

\section*{Acknowledgements}
In this work we made use of asteroid physical properties data from \textit{mp3c.oca.eu, Observatoire de la C\^ote d'Azur}, whose database is also mirrored at \textit{https://www.cosmos.esa.int/web/astphys}. We thank J. Hanus for his help on asteroid masses and sizes and A. Morbidelli for his help on the collisional model. We would like to thank the anonymous reviewer for his/her comments and suggestions, which led to significant improvement of the manuscript. This work was partially supported by the French National Research Agency under the project ``Investissements d'Avenir'' UCA$^{JEDI}$ with the reference number ANR-15-IDEX-01.
\bibliographystyle{mnras}
\bibliography{references.bib} 



\bsp	
\label{lastpage}
\end{document}